\documentclass[prb,twocolumn]{revtex4}
\pdfoutput=1

\usepackage{bm}
\usepackage{graphicx} 
\usepackage{notes2bib}

\begin{document}

\title{HELICITY SENSITIVE PLASMONIC TERAHERTZ INTERFEROMETER}

\author{Y. Matyushkin,$^{1,2,5}$ S. Danilov,$^3$ M. Moskotin,$^{1,2}$ V. Belosevich,$^{1,2}$  N. Kaurova,$^2$ \\ M. Rybin,$^1$ E. Obraztsova,$^1$  G. Fedorov,$^{1,2}$   I. Gorbenko,$^4$  V. Kachorovskii,$^4$ and S. Ganichev$^3$}
\affiliation{$^1$Moscow Institute of Physics and Technology (State University), Dolgoprudny, Russia} 
\affiliation{$^2$Physics Department, Moscow State University of Education (MSPU), Moscow, Russia} 
\affiliation{$^3$Terahertz Center, University of Regensburg, Regensburg, Germany} 
\affiliation{$^4$Ioffe Institute, St. Petersburg, Russia}
\affiliation{$^5$National Research University Higher School of Economics, Moscow , Russia} 
		
\begin{abstract}
Plasmonic interferometry is a rapidly growing area of research with a huge potential for applications in terahertz frequency range. In this Letter, we explore a   plasmonic interferometer based on graphene Field Effect Transistor connected to  specially designed antennas. As a key result, we observe helicity- and phase-sensitive conversion of circularly-polarized  radiation into dc photovoltage caused by the plasmon-interference  mechanism: two plasma waves, excited at the source and drain part of the  transistor interfere inside the channel. The helicity sensitive phase shift  between these waves is achieved by using an asymmetric antenna configuration. The dc signal  changes sign with inversion of   the helicity.  Suggested plasmonic interferometer is  capable for measuring of phase difference between  two arbitrary phase-shifted optical signals. 		The observed effect opens a wide avenue for phase-sensisitve  probing of plasma wave excitations in two-dimensional materials.
\end{abstract}

\maketitle

	\section{Introduction}

Interference is in heart of quantum physics and classical optics, where wave superposition plays a key role \cite{Scully1997, Hariharan2007, DemkowiczDobrzaski2015}. Besides fundamental significance, interference has very important applied aspects. Optical and electronic interferometers are actively used in modern electronics, and the range of applications is extremely wide and continuously expanding. In addition to standard applications in optics and electronics \cite{Scully1997, Hariharan2007, DemkowiczDobrzaski2015}, exciting examples include multiphoton entanglement \cite{Pan2012}, nonperturbative multiphonon interference~\cite{ch3Ganichev86p729,ch3Keay95p4098}, atomic and molecular interferometry \cite{Boal1990,Cronin2009, Uzan2020} with recent results  in cold-atoms-based precision interferometry \cite{Becker2018}, neutron interferometry \cite{Danner2020}, interferometers for medical purposes \cite{Yin2019}, interference analysis of turbulent states \cite{Spahr2019}, qubit interferometry \cite{Shevchenko2010} with a recent analysis of the Majorana qubits \cite{Wang2018}, numerous amazing applications in the astronomy \cite{Saha2002, Goda2008, Adhikari2014, Sala2019}, such as interferometers for measuring of gravitational waves \cite{Goda2008, Adhikari2014} and antimatter wave interferometry \cite{Sala2019}, etc.
	
	Recently, a new direction, plasmonic inter\-fe\-ro\-metry \cite{Gramotnev2010, Graydon2012, Ali2018, Khajemiri2018, Khajemiri2019, Zhang2015, Salamin2019, Yuan2019, Woessner2017, Smolyaninov2019, Hakala2018, Dennis2015, Haffner2015}, has started to actively develop. The plasma wave velocity in 2D materials is normally an order of magnitude larger than the  electron  drift velocity and is much smaller than the speed of light. Hence, the plasmonic  submicron  sized interferometers based on 2D materials are expected to operate efficiently in the terahertz (THz) frequency range \cite{mittleman_2010, Dhillon2017}. In particular, it has been predicted theoretically that a field-effect transistor (FET) can serve as a simple device for studying plasmonic interference effects~\cite{Drexler2012,Romanov2013, gorbenko2018single, Gorbenko2019}. Specifically, it was suggested that a FET with two antennas attached to the drain and source shows a dc current response to circularly polarized THz radiation which is partially driven by the interference of plasma waves and hence by helicity of incoming radiation.  First experimental hint on existence of such an interference contribution was reported in Ref.~[$\!\!$\citenum{Drexler2012}] for an industrial FET, where helicity-driven effects were obtained due to  unintentional peculiarities of contact pads. Despite the first successes, creation of effective plasmonic interferometers  is still   a  challenging task although  in many aspects plasmonic-related THz phenomena are sufficiently well studied  \cite {Dyakonov1993, Dyakonov1996, Knap2009, Tauk2006, Sakowicz2008, JOAP_KnapKachorovskii2002, APL_Knap2002, Peralta2002, Otsuji2004, TeppeKnap2005, Teppe2005, Veksler2006, Derkacs2006, ElFatimy2006}   with some commercial applications already in the market. Appearance of graphene opened rout for a novel class of active plasmonic structures   \cite{Vicarelli2012} promising for plasmonic inteferometry due to non-parabolic dispersion of charge carriers and support of weakly decaying plasmonic excitations~\cite{Koppens2014}.
	
	In this Letter, we explore an all-electric tunable -- by the gate voltage -- plasmonic interferometer based on graphene FET  connected to specially designed antennas. Our interferometer demonstrates helicity-driven conversion of incoming circularly-polarized radiation into  phase- and helicity-sensitive dc photovoltage signal. 	The effect is detected at room- and liquid helium- temperatures for radiation frequencies in the range from 0.69 to 2.54~THz. All our results show plasmonic nature of effect. Specifically, the rectification of the interfering plasma waves leads to dc response, which is controlled by the gate voltage and encodes information about helicity of the radiation and phase difference between the plasmonic signals. A remarkable feature of this plasmonic interferometer is that there is no need to create an optical delay line, which has to be comparable with the quite large wavelength of the THz signal. By contrast, in this setup, phase shift between the plasma waves excited at the source and drain electrodes of the FET is maintained by combination of the antenna geometry and the radiation helicity. It remains finite even in the limit of infinite wavelength and changes sign with inversion of the radiation helicity. The  plasmonic interferometer concept realized in our work opens a wide avenue for phase-sensitive probing of plasma wave excitations in two-dimensional materials.

\begin{figure*}
		\includegraphics[width=\linewidth]{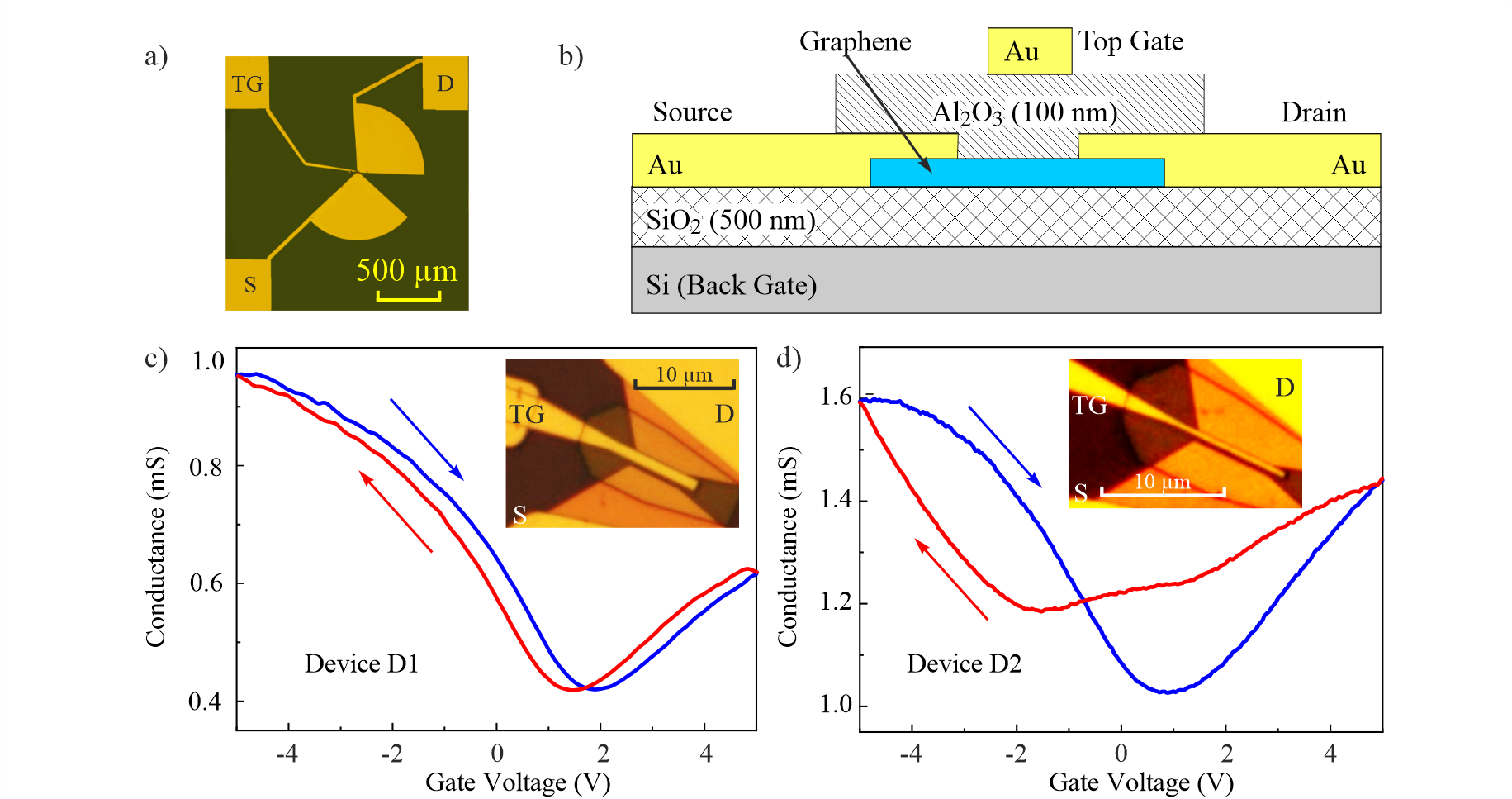}
		\caption{Devices configuration and characterization. (a) Optical image illustrating the device layout with source and drain electrodes connected to sleeves of a bent bow-tie antenna. (b) Structures cross-section showing relative location of the source, drain and top gate electrodes as well as  thickness of the dielectric layers. Panels (c) and (d) transfer characteristics of devices 1 and 2, respectively. The curves are measured at a bias voltage of 10~mV. The data are presented for two directions of the gate voltage sweeps, which yield different positions of the charge neutrality point. Insets show zoomed images of the devices.}
		\label{fig:characterization}
	\end{figure*}

	\section{Devices and measurements}
	The single layer graphene (SLG), acting as the conducting channel of a field-effect transistor, was synthesized in a home-made cold-wall chemical vapor deposition reactor by chemical vapor deposition (CVD) on a copper foil with a thickness of 25~$\mu$m \cite{Rybin2016}. SLG was transferred onto an oxidized silicon wafer\cite{Gayduchenko2018}, for more details see Appendix.
	The antenna sleeves were attached to the source and drain electrodes. To realize the helicity sensitive terahertz plasmonic interferometer, the antenna sleeves were bent by 45${}^\circ$ as shown in Fig.~\ref{fig:characterization}b. The sleeves were made using photolithographic methods and metallization sputtering (Ti/Au, 5/100~nm). 
	At the first lithography step, contact pads to graphene channel were formed using pure Au with a thickness of 25~nm. Note that we did not use adhesion sub layer (like titan or chrome) at this step.	At the next technological step, e-beam evaporator was applied to sputter a 100~nm thick layer of Al$_2$O$_3$ that acts as a gate insulator. Note that Al$_2$O$_3$ layer reduces the initially high doping level of as-transferred graphene to almost zero \cite{Fallahazad2012, Kang2013}. 
	Finally, the top gate electrode (Ti/Au, 5/200~nm) is patterned and formed using standard sputtering and lift-off techniques. The resulting structure is sketched in the Figure~\ref{fig:characterization}a. Two devices with channel lengths 2~$\mu$m (device~1) and 1~$\mu$m (device~2) were fabricated. Zoomed images of the channel parts are shown in insets in Figs.~\ref{fig:characterization}c and \ref{fig:characterization}d. Note that for both devices the gates are deposited asymmetrically in respect to the channel. They cover about 75\% (device~1) and 50\% (device~2) of the channels and the gate stripes are located closer to the drain contact pads. 
	
	Typical transport characteristics of our graphene devices are shown in the Figs.~\ref{fig:characterization}c and \ref{fig:characterization}d. For different directions of the gate voltage sweep as well as the sample cooldowns, the charge neutrality point (CNP) can occur at different gate voltages $U_{\rm g}$. This is well known feature and is possibly caused by the gate-sweep direction (cooldown-) dependent charge trapping. Therefore, below we indicate range of $U_{\rm g}$ corresponding to the CNP instead of providing its exact value. Comparing conductance curves of the device~1 and 2 shows that the minimal conductance $G$ scales as an inverse of the graphene channel length, indicating that most of the resistance comes from the graphene channel itself rather than from the contacts. Analysis of the channel mobilities and corresponding scattering times is presented in Appendix. 
	Using the Drude formula generalized for graphene we estimate scattering times of the order of 10-20~fs for, e.g., device~1 at room temperature. Finally, we note that the $G(U_{\rm g})$ curves do not change as the temperature goes down (see Appendix), meaning that mobility is restricted by the defect scattering. 
	
	The experiments have been performed applying a continuous wave methanol laser operating at frequencies $f_1 = 2.54$~THz (wavelength $\lambda_1 = 118$~$\mu$m) with a power of $P \approx 20$~mW and $f_2 = 0.69$~THz (wavelength $\lambda_2 = 432$~$\mu$m) with $P \approx 2$~mW \cite{Ganichev2009, Kvon2008}. 
	The laser spot with a diameter of about 1-3~mm is substantially larger than the sample size ensuring uniform illumination of both antennas.
	The radiation polarization state was controllably varied by means of lambda-half plate that rotates the polarization direction of linear polarized radiation and lambda-quarter plate that transforms linearly polarized radiation into elleptically polarized one.
		
	The helicity of the radiation is then controlled via changing the angle $\phi$ between the laser polarization and the main axes of the lambda-quarter plate, so that for $\phi = 45^\circ$ the radiation is right circularly polarized ($\sigma^+$) and for $\phi = 135^\circ$ - left circularly polarized ($\sigma^-$). The functional behavior of the Stokes parameters upon rotation of the waveplates is summarized in Appendix, see also Ref.~[$\!\!$\citenum{Belkov2005}]. 
	The samples were placed in an temperature variable optical cryostat and photoresponse was measured as the voltage drop $U$ directly over the sample 
	applying lock-in technique at a modulation frequency of 75~Hz.
	
	\begin{figure*}
		\includegraphics[width=\linewidth]{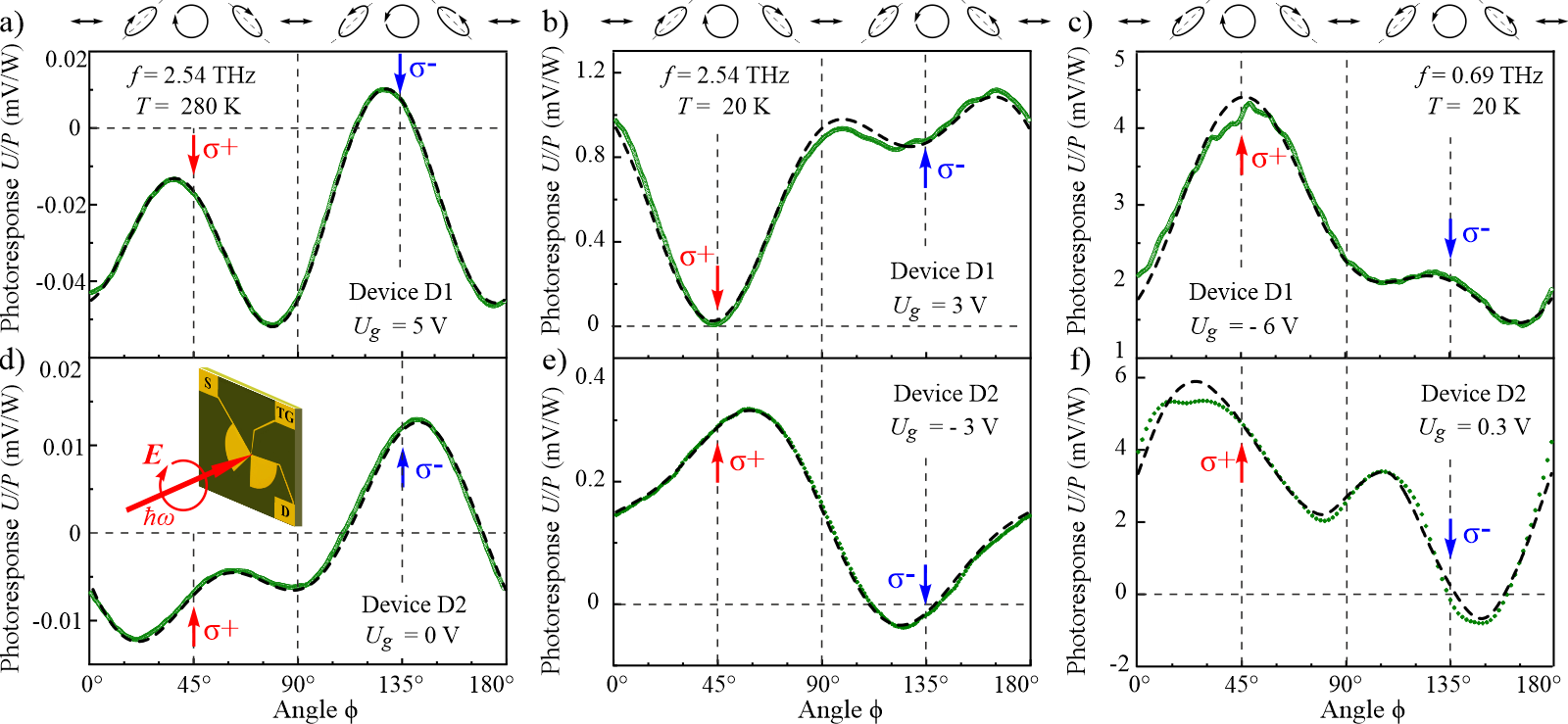}
		\caption{Helicity dependence of the photovoltage $U(\phi)$  normalized by the radiation power $P$. Upper panels (a--c) show the results obtained in device~1 and lower panels (d--f) those in device~2. The data are shown for two radiation frequencies ($f = 2.54$ and 0.69~THz), two temperatures (room temperature and $T = 20$~K) and different gate voltages $U_{\rm g}$.  Dashed lines show fits according to Eq.~(\ref{eqn:eq1}). The values of the fitting parameters $U_{\rm C}, U_{\rm L1}, U_{\rm L2},$ and $ U_0$ are given in Appendix. The ellipses on top illustrate the polarization states at different angles $\phi$. The inset sketches the experimental geometry.
		}
		\label{fig:helicity}
	\end{figure*}
	
	\section{Results}
	The principal observation made in our experiment is that for all investigated devices the response to a circularly polarized radiation 
	crucially depends on its helicity. Fig.~\ref{fig:helicity} displays the response voltage $U$ normalized by the radiation intensity as a function of the angle $\phi$ obtained under different conditions. 
	We emphasize significant difference in the signal for $\phi = 45^\circ$ and $135^\circ$, corresponding to opposite helicities of circularly polarized light, in particular, the sign inversion observed under some conditions, see e.g. Fig.~\ref{fig:helicity}d.
	The effect is observed for radiation with frequencies 2.54 and 0.69~THz in a wide temperature range from 4.2~K to 300~K. The overall dependence of the signal on angle $\phi$ is more complex and is well described by
	\begin{eqnarray} \label{eqn:eq1}
	U(\phi)  &=& U_{\rm C}\sin(2\phi)+U_{\rm L1}\sin(4\phi)/2+
	\\ \nonumber
	&+&U_{\rm L2}[\cos(4 \phi)+1]/2 + U_0
	\end{eqnarray}
	with $U_{\rm  C} $, $U_{\rm L1}$, $U_{ \rm L2}$  and $U_0$ are fit parameters  depending on gate voltage, temperature and radiation frequency. Note that trigonometric functions used for the fit are the radiation Stokes parameters describing the degree of the circular and linear polarization (see Appendix)
	\cite{bahaasaleh2019,Belkov2005, Weber2008}. While three last terms are insensitive to the radiation helicity the first term ($\propto U_{\rm C}$) is $\pi$-periodic and describes helicity-sensitive response: it reverses the sign upon switching from right- ($\sigma^+$) to left- ($\sigma^-$) handed circular polarization. Figure~\ref{fig:helicity} reveals that this term gives substantial contribution to the total signal.
	As we show below the $\pi$-periodic term is related to the plasma interference in the graphene-based FET channel. Measurements at room and low temperatures demonstrate that cooling the device increases the amplitude of the circular photoresponse by more than ten times, see Figs.~\ref{fig:helicity}a and \ref{fig:helicity}b as well as \ref{fig:helicity}d and \ref{fig:helicity}e. The signal increase is also observed by reduction of radiation frequency, see Figs.~\ref{fig:helicity}b and \ref{fig:helicity}c as well as \ref{fig:helicity}e and \ref{fig:helicity}f.
	
	\begin{figure*}
		\includegraphics[width=\linewidth]{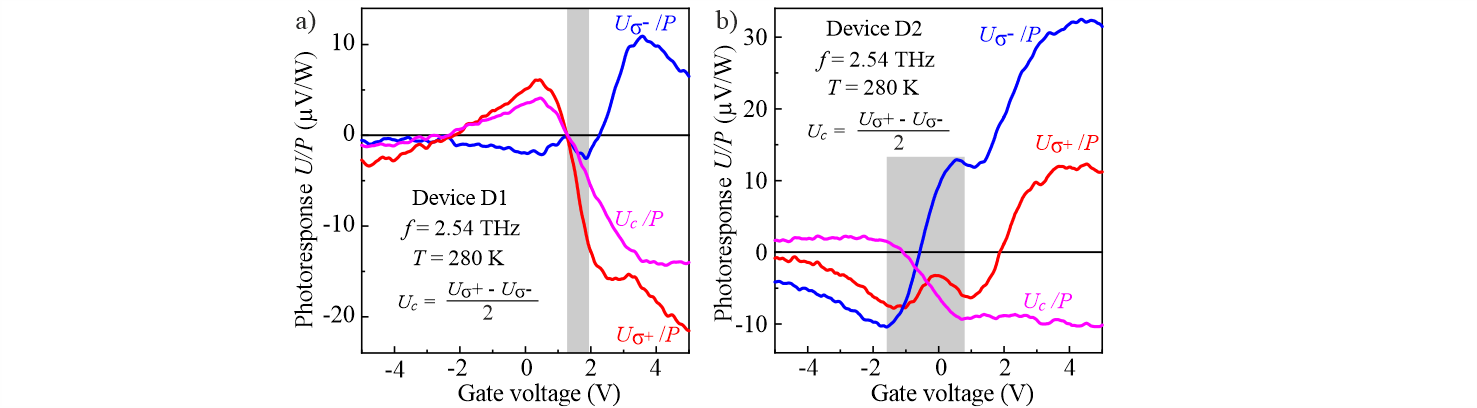}
		\caption{Gate voltage dependencies of the photoresponse of the devices 1 (panel a) and 2 (panel b).  Red and blue curves show responses to right- ($U_{\sigma^+}$) and left- ($U_{\sigma^-}$) handed circularly polarized radiation, respectively. Magenta curve shows the amplitude of the helicity driven response $U_{\rm C} = (U_{\sigma^+}-U_{\sigma^-})/2$.
			Shadowed areas show the range of CNP obtained by transport measurements with different sweeps of the gate voltage $U_{\rm g}$ , see Figs.~\ref{fig:characterization}c and d.}
		\label{fig:responses}
	\end{figure*}
	
	Having experimentally proved the applicability of Eq.~(\ref{eqn:eq1}) and substantial contribution of the helicity drive signal we now concentrate on the dependence of the parameter $U_{\rm C}$ on the gate voltage that 
	controls the type and concentration of the charge carriers in the FET channel. 
	We use the following procedure: we obtain the gate voltage dependence of the response voltage normalized to the radiation power $P$ for two positions of the $\lambda/4$ plate $\phi = 45^\circ$ and $135^\circ$, corresponding to opposite helicities of circularly polarized light ($\sigma^+$ and $\sigma^-$). 
	The half-difference between these two curves directly gives gate voltage dependence of the helicity sensitive photoresponse 
	$U_{\rm C} = (U_{\sigma^+} - U_{\sigma^-})/2\,.$
	[see Eq.~(\ref{eqn:eq1})]
	Results of these measurements, shown in Fig.~\ref{fig:responses}, reveal that $U_{\rm C}$ is more pronounced at positive gate voltage, where the channel is electrostatically doped with electrons, and changes the sign close to the CNP. As addressed above the variation of the CNP from measurement to measurement does not allow us to allocate the exact position of the CNP for the gate voltage 
	sweeps during the photoresponse measurements.
	Note that for the device~1, having the gate length twice larger as that of device~2, at large negative gate voltages the second sign inversion of the photocurrent is present. Figure~\ref{fig:responses}a indicates that in device~1 for the whole range of gate voltages photoresponse for $\sigma^+$- and $\sigma^-$- radiation have consistently opposite sign indicating negligible contribution of the polarization independent background. In device~2, however, the background is of the same order as the helicity sensitive response $U_{\rm C}$, see Fig.~\ref{fig:responses}b.
	
	\begin{figure*}
		\includegraphics[width=\linewidth]{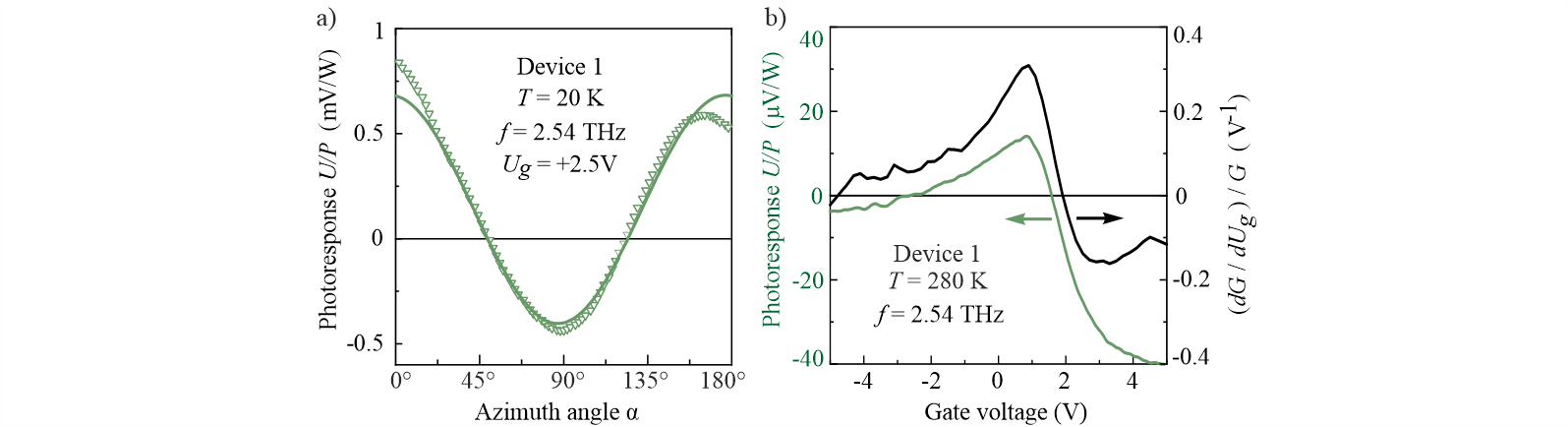}
		\caption{Photovoltage as a function of the azimuth angle $\alpha$. The data are obtained applying linearly polarized radiation with $f = 2.54$~THz for device 1 at $T = 20$~K. Solid line shows fit after   Eq.~(\ref{eqn:L1L2}) with fitting parameters: $U_{\rm L1}/P = -0.16$, $U_{\rm L2}/P = 0.54$ and $U_0/P = 0.14$~mV/W. Panel (b) shows gate voltage dependence obtained for device 1 at $T=280$~K applying linearly polarized radiation ($\alpha=0$) with frequency $f = 2.54$~THz (green line). Black line shows gate voltage dependence of the normalized first derivative of the conductance $G$ over $U_{\rm g}$: $(dG/dU_{\rm g})/G$.}
		\label{fig:azimuth}
	\end{figure*}
	
	Finally, we present additional data on the contributions proportional to fit parameters $U_{\rm L1}$, $U_{\rm L2}$ and $U_0$ in Eq.~(\ref{eqn:eq1}). As discussed above these contributions do not connected to the radiation helicity and are related to the Stokes parameters describing the degrees of linear polarization (terms $\propto U_{\rm L1}$ and $\propto U_{\rm L2}$) and radiation intensity (term $\propto U_0$). In experiments applying linearly polarized radiation with rotation of $\lambda/2$ plates Stokes parameters modifies and polarization dependence Eq.~(\ref{eqn:eq1}) takes a form
	\begin{equation}
	U(\alpha) = U_{\rm L1}\sin(2\alpha) + U_{\rm L2}\cos(2\alpha) + U_0 .
	\label{eqn:L1L2}
	\end{equation}
	An example of the photoresponse variation upon change of the azimuth angle $\alpha$ is shown in Fig.~\ref{fig:azimuth}a. The data reflects the specific antenna pattern of our devices with tilted sleeves. Figure~\ref{fig:azimuth}b shows the gate voltage dependence of the photoresponse obtained in device~1 for $\alpha=0$. Comparing these plots with the results for circular photoresponse shows that they behave similarly: in both cases signal changes the sign close to CNP and the response for positive gate voltages is larger than that for negative $U_{\rm g}$. Transport measurements carried out parallel to 
	photoresponse measurements show that the photosignal behaves similarly to the normalized first derivative of the conductance $G$ over $U_{\rm g}$: $(dG/dU_{\rm g})/G$, see Fig.~\ref{fig:azimuth}a. Note that this behavior is well known for non-coherent, phase-insensitive plasmonic detectors \cite{Veksler2006}.
	
	\section{Theory and discussion}
	Conversion of THz radiation into dc voltage can be obtained	due to 
	several phenomena including photothermoelectric (PTE) effects~\cite{Fedorov2013, Gabor2011, Fuhrer2014}, rectification on inhomogeneity of carrier doping in gated structures \cite{Fuhrer2014, Gayduchenko2018, CastillaKoppens2019}, photogalvanic and photon drag effects \cite{Karch2010, Jiang2011, Glazov2014} as well as rectification of electromagnetic waves in a FET channel supporting plasma waves \cite{Dyakonov1996}. 
	However, in our experiment only plasmonic mechanism can yield the dc voltage whose polarity changes upon switching the radiation helicity. Indeed, PTE effects and rectification due to
	gradient of carrier doping in gated structures are based on inhomogeneities of either radiation heating or radiation absorption, which are helicity-insensitive~\cite{b1}.
	%
	
	Below, we show that the  helicity-sensitive plasmonic response originates from the interference of plasmonic signals excited by the source and drain antenna sleeves. 
	The source and drain potentials with respect to the top gate are given by
	\begin{equation}
	U_{\rm A,B} (t) = U_{\rm A,B}\cos(\omega t- \varphi_{\rm A,B}),
	\label{eqn:time_dep}
	\end{equation}
	where $\omega$ is the radiation frequency.
	Complex amplitudes $U_{\rm A} e^{i \varphi_{\rm A}}$ and $U_{\rm B} e^{i \varphi_{\rm B}}$ of the signal at source and drain, respectively,  are proportional to the electric field amplitude of the incident electromagnetic wave. 
	Their amplitudes and the phase shift  between signals ($\varphi_{\rm A}-\varphi_{\rm B}$) depend on the radiation polarization and antennas geometry, see Appendix.  Design of our devices, see Fig.~\ref{fig:interpretation}a, ensures asymmetric coupling of radiation to the source and drain electrodes so that both amplitudes and phases of source and drain potentials are different. 
	Most importantly, when such a bent bow-tie antenna is illuminated by circularly polarized radiation, the source- and drain-related antenna sleeves are polarized with a time delay because of rotation of the electric field vector. 
	For circularly polarized wave, the phase shift  changes sign with changing the helicity of the radiation: $ \varphi_{\rm A}-\varphi_{\rm B}=\theta_{\rm A}-\theta_{\rm B}, $ for 
	$\omega>0$ (positive helicity)
	and  $ \varphi_{\rm A}-\varphi_{\rm B}=-(\theta_{\rm A}-\theta_{\rm B}), $ for 
	$\omega<0$ (negative helicity ),  where $\theta_A$ and $\theta_B$ are  geometrical angles of antennas sleeves. Equations describing more general case of elliptic polarization are derived in Appendix. We note that in the case of linear polarization, the phase shift is zero and the second term in Eq.~(\ref{eqn:response1}) is absent.

	\begin{figure*}
		\includegraphics[width=\linewidth]{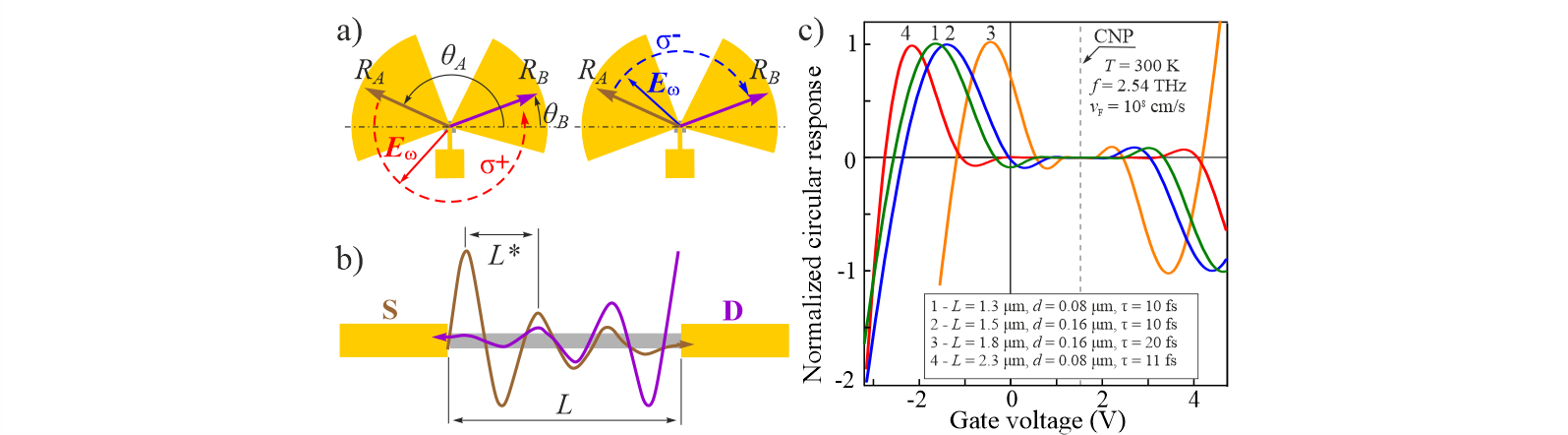}
		\caption{ Panels (a) and (b) illustrate the physics behind the circular photoresponse caused by the plasmonic interference. (a) Bent bow-tie antenna characterized with two vectors $R_{\rm A}$ and $R_{\rm B}$ along with the hodograph of the electric filed in case of circularly polarized for positive (left, red arrow) and negative (right, blue arrow) helicities. Due to opposite rotation direction the phase differences between the source and drain potentials have opposite signs for  opposite helicities. (b) Illustration of plasma waves  excited at the source and drain electrodes. (c)  Calculated gate dependence of the interference part of the response. Calculations are done for different  devices with parameters given in the plot labels. Vertical dashed line shows the position of the charge neutrality point.}
		\label{fig:interpretation}
	\end{figure*}
	
	Due to hydrodynamic non-linearity of plasma waves~\cite{Dyakonov1993,Dyakonov1996}  DC voltage  across  the channel  appears 
	\begin{equation}
	U = F_1(U_{\rm A}^2-U_{\rm B}^2)+F_2U_{\rm A}U_{\rm B}\cdot \sin (\varphi_{\rm A} - \varphi_{\rm B}),
	\label{eqn:response1}
	\end{equation}
	where $F_1$ and $F_2$  are gate-controlled coefficients which represent, respectively,  non-coherent\cite{Dyakonov1996} and  interference \cite{Drexler2012,Romanov2013,gorbenko2018single,Gorbenko2019}  contributions to the response. 
	This coefficients do not depend on signal phases and amplitudes, while all information about coupling with antennas is encoded in factors $(U_{\rm A}^2-U_{\rm B}^2)$ and $U_{\rm A}U_{\rm B}\sin(\varphi_{\rm A} - \varphi_{\rm B})$.
	
	For the case of radiation with arbitrary polarization characterized by  the Stockes parameters, that are controlled by the orientation of the $\lambda/4$ plate, defined by the phase angle $\phi$, we obtain   
	\begin{eqnarray} \nonumber
	U(\phi)&= 
	F_1 \left[a_0 + a_{\rm L1} \frac{\sin (4 \phi)}{2}+ a_{\rm L2}\frac{1+ \cos(4\phi)}{4}
	\right] +    \nonumber    \\ &+
	F_2~ a_{\rm C}  \sin(2 \phi).
	\label{eqn:response2}
	\end{eqnarray}
	Here  $a_0$, $a_{\rm L1}$, $ a_{\rm L2}$, $a_{\rm C}$ are the geometrical  factors calculated in Appendix for a simplified model. 
	Note that these factors are independent of the gate voltage, so that the entire gate voltage dependence of the photoresponse is defined by the factors $F_1$ and $F_2$.  Comparing  these results with empirical Eq.~(\ref{eqn:eq1}) we conclude that the data shown in the Fig.~\ref{fig:helicity} are fully consistent with theoretical considerations discussed thus far. In particular, the interference-induced helicity-sensitive contribution, given by the last term on the right hand side of the above equation, is clearly observed in the experiment, see Fig.~\ref{fig:helicity}. 
	
	Such interference contribution appears when source and drain electrodes ``talk'' to each other via exchange of plasma wave phase-shifted  excitations. Hence, the  characteristic length of plasma wave decay $L_*$   should not be  too small  as compared to channel length $L$ so that plasmons excited near source and drain electrodes could efficiently interfere within the channel, see the Fig.~\ref{fig:interpretation}b. 
	As the gate voltage controls the type and concentration of the charge carriers it also controls the  velocity of the plasma waves $s$  and the length $L_*.$ 
	As a result, the second term in the Eq.~(\ref{eqn:response1}) should oscillate as function of the gate voltage. 
	The general formulas for response given in Appendix can be  essentially simplified for the non-resonant case $ s/L \ll \gamma, ~  \omega \ll \gamma$,  which corresponds to our experimental situation~\cite{b2}.
	%
	%
	In this case, plasma waves decay from the source and drain part of the channel within the  length
	$   L_*= {s\sqrt 2}/{\sqrt {\omega \gamma}},$
	and the response is given by Eq.~(\ref{F_2}) in Appendix. The parameters $F_{1,2}$ in   Eq.~(\ref{eqn:response2})  are given by
	\begin{equation}
	F_1=\frac{1}{4U_{\rm g}}, \quad F_2= \frac{4\omega }{\gamma}~ \frac{ \sin \left({L}/{L_*} \right) e^{-L/L_*}}{U_{\rm g}}.      
	\label{F12}
	\end{equation}
	In our experiment $L_*$ was  essentially smaller  than the device length: $L_*  \sim (0.1 \div 0.3) L $.  However, the interference, helicity-dependent part of the response was clearly seen in the experiment.  The result of calculations 
	are presented in Fig.~\ref{fig:interpretation}c. For gate voltages far from the Dirac point we obtain a qualitative agreement of the calculations and results of experiments presented in Fig.~\ref{fig:responses}a: 
	\begin{itemize}
		\item the circular photoresponse at high positive gate voltages and for moderate negative $U_{\rm g}$ have opposite sign;
		\item with  increase of the negative gate voltages value the response changes its sign;
		\item in the vicinity of the Dirac point calculations yield oscillations of the circular response.
	\end{itemize}
	
	The last statement needs a clarification.   While the oscillations are not visible in the experimental magenta  curve of Fig.~\ref{fig:responses}a, showing $U_{\rm C} = (U_{\sigma^+} - U_{\sigma^-})/2$, in individual curves obtained for left- (blue curve) and right- (red curve) they are clearly present. 
	This difference is caused by the fact that the $U_{\sigma^+}$  and  $U_{\sigma^-}$ curves  represent  the results of two different experiments, namely,  measurements for $\sigma^+$ and $\sigma^-$ radiation. At the same time, $U_{\rm C}$ is obtained as a result of subtraction  of these two curves, corresponding to different $U_{\rm g}$ sweeps. Due to the  hysteresis of the $R_{xx}$ discussed above in Sec. 2,  the sample parameters were slightly different for these two measurements.  As a result,  the oscillations present in one curve are superimposed with larger featureless signal from the other. 
	Furthermore, Fig.~\ref{fig:responses}b shows that presence of the oscillations in photoresponse to circularly polarized radiation for the second device too.  
	
	Now we estimate the period of the oscillations. The dependence on the gate voltage is mostly encoded in ${L_* \propto {U_{\rm g}}^{1/4} }.$  The oscillations period can be estimated from the condition ${\delta (L/L_*) \sim 1,}$ which gives  ${(L/L_*)\delta U_{\rm g}/4 U_{\rm g}\sim 1.}$  For ${U_{\rm g} \approx 2}$ V and    ${L_*/L \approx 0.1,}$  we find ${\delta U_{\rm g} \approx 0.8}$~V in a good agreement with experiment. We also note  that the experimentally observed oscillations (see blue curve for the device 1) decays at the same scale as an oscillation period in an excellent  agreement  with behavior of the function $F_2$, see  Eq.~(\ref{F12}).  
	
	Finally, we emphasize  that the presence of oscillations in the   hallmark  of the interference part of response.  Importantly, the  response to the linearly polarized radiation does not show any oscillations in the vicinity of the CNP, see Fig.~\ref{fig:azimuth}b. By contrast it just follows to $(dG/dU_{\rm g})/G$ --- a well know behavior for Dyakonov-Shur non-coherent plasmonic  detectors, see e.g. Ref.~[$\!\!$\citenum{Veksler2006}]. Note that for linearly polarized radiation the above theory also yields this gate dependence, see structureless expression for  non-coherent contribution  $F_1$  in  Eq.~(\ref{F12}).
	
	\section{Summary}
	To summarize, we demonstrated that specially designed graphene-based FET can be used to study plasma wave interference effects. Our approaches can be extrapolated to other 2D materials and used as a tool to characterize optically-induced plasmonic excitations. Specifically, the 
	conversion of the interfering plasma waves into
	dc response	is controlled by the gate voltage and encodes information about helicity of the radiation and phase difference between the plasmonic signals.
	Remarkably,  our work shows that CVD graphene with moderate mobility, which is compatible with most standard technological routes can be used as a material for active plasmonic devices. 
	The suggested device design can be used for a broad-band helicity-sensitive interferometer capable of analyzing both polarization of THz radiation and geometrical phase shift caused by antennas asymmetry. 
	By the proper choice of the antenna design and  FET parameters, 
	phase-sensitive and fast room temperature plasmonic detectors can be tuned to detect individual Stokes parameters.
	Hence, our work  paves a novel way for developing the all-electric detectors of the terahertz radiation polarization state.

	\section{Acknowledgements}
		The work was supported by the FLAG-ERA program (project DeMeGRaS, project GA501/16-1 of the DFG), the Russian Foundation for Basic Research within Grants No. 18-37-20058 and No. 18-29-20116, the Volkswagen Stiftung Program (97738) and Foundation for Polish Science (IRA Program, grant MAB/2018/9, CENTERA).  The work of V.K. was supported by  the Russian Science Foundation (Grant No. 20-12-00147). The work of I.G. was   supported by Russian Foundation for Basic Research (Grant No. 20-02-00490)  and  by Foundation for the Advancement of Theoretical Physics and Mathematics BASIS. 
		M.M. acknowledges support of Russian Science Foundation (project No. 17-72-30036) in sample design.

	
	\section{Appendix}
	\subsection{Devices, experimental details and fit parameters}

\textbf{Device fabrication.} The single layer graphene (SLG), acting as the conducting channel of a field-effect transistor, was synthesized in a home-made cold-wall chemical vapor deposition reactor by chemical vapor deposition (CVD) on a copper foil with a thickness of 25~$\mu$m \cite{Rybin2016}. SLG was transferred onto an oxidized silicon wafer\cite{Gayduchenko2018}. The silicon substrate consists of 480~$\mu$m thick silicon wafer covered with a 500~nm thick thermally grown SiO$_2$ layer. Note that silicon used for the substrate (with the room temperature resistivity of 10~$\Omega\cdot$cm) is transparent for the sub-THz and THz radiation. After transferring  graphene onto a silicon wafer the geometry of the device is further defined by e-beam lithography using a PMMA mask and oxygen plasma etching. 

At the first lithography step, contact pads to graphene channel were formed using pure Au with a thickness of 25~nm. Note that we did not use adhesion sub layer (like titan or chrome) at this step.	At the next technological step, e-beam evaporator was applied to sputter a 100~nm thick layer of Al$_2$O$_3$ that acts as a gate insulator. Note that Al$_2$O$_3$ layer reduces the initially high doping level of as-transferred graphene to almost zero \cite{Fallahazad2012, Kang2013}. 

Afterward, the antenna sleeves were attached to the source and drain electrodes. To realize the helicity sensitive terahertz plasmonic interferometer, the antenna sleeves were bent by  45${}^\circ$ as shown in Fig.~\ref{fig:characterization}a. The sleeves were made using photolithographic methods and metallization sputtering (Ti/Au, 5/100~nm). Finally, the top gate electrode (Ti/Au, 5/200~nm) is patterned and formed using standard sputtering and lift-off techniques. The resulting structure is sketched in  Fig.~\ref{fig:characterization}b of the main text. Two devices with channel lengths 2~$\mu$m (device~1) and 1~$\mu$m (device~2) were fabricated. Zoomed images of the channel parts are shown in insets in Figs.~\ref{fig:characterization}c and d of the main text. Note that for both devices the gates are deposited asymmetrically in respect to the channel. They cover about 75\% (device~1) and 50\% (device~2) of the channels and the gate stripes are located closer to the drain contact pads. 

\textbf{Transport characteristics and transport scattering time.} Typical transport characteristics of our graphene devices are shown in the Figs.~1c and 1d of the main text. For different directions of the gate voltage scan as well as the sample cooldowns, the charge neutrality point (CNP) can occur at different gate voltages. This is well known feature and is possibly caused by the gate-scan-direction (cooldown-) dependent charge trapping. Therefore, below we indicate range of $U_{\rm g}$ corresponding to the CNP instead of providing its exact value. Comparing conductance curves of the device~1 and 2 shows that the minimal conductance $G$ scales as an inverse of the graphene channel length, indicating that most of the resistance comes from the graphene channel itself rather than from the contacts.  For that we used the transfer curves measured in two-probe configuration, since we cannot perform four contacts Hall effect measurements due to configuration of our devices.

\begin{figure}[h!]
	\centerline{\includegraphics[width=\columnwidth]{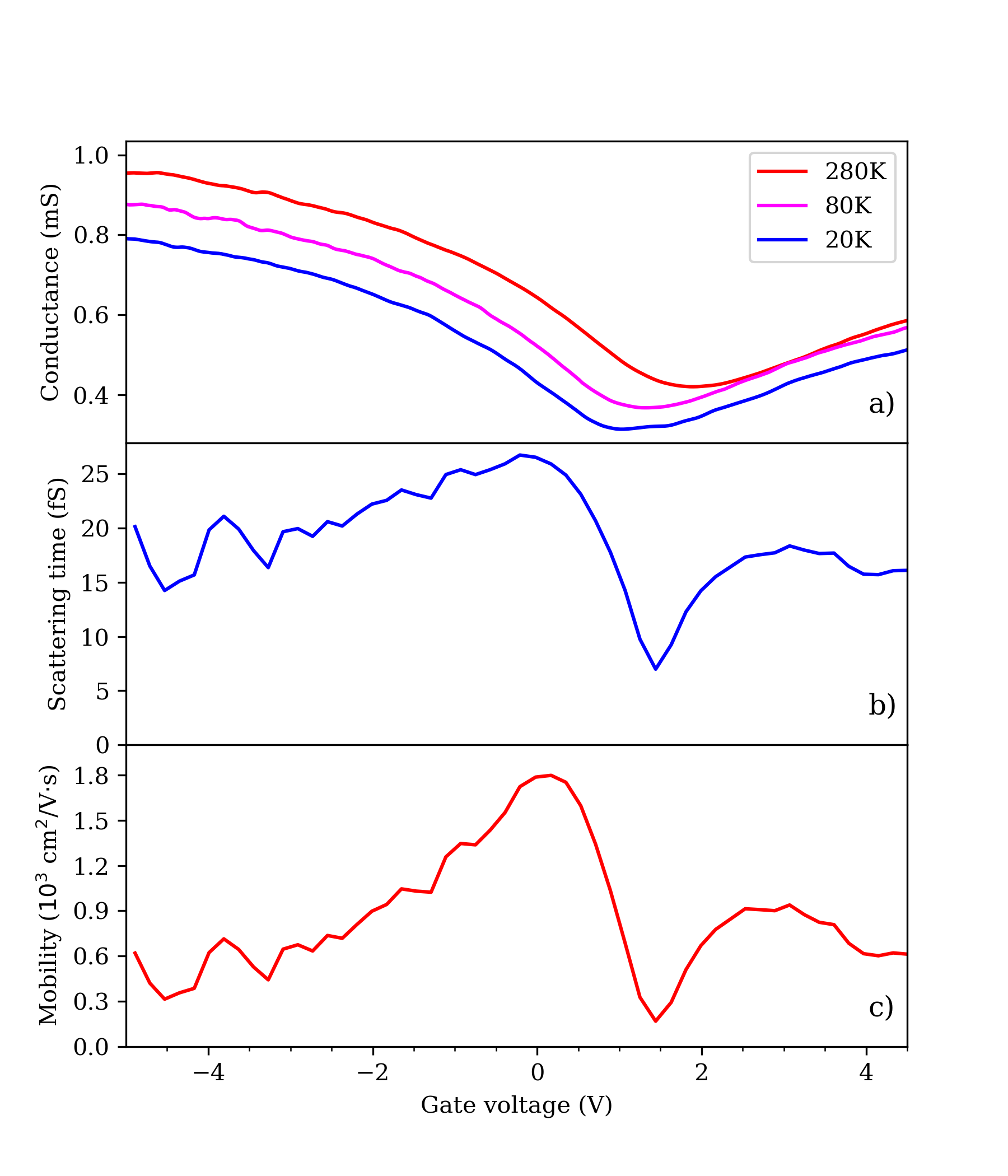}	} 
	\caption{(Color online) Electron transport in the our CVD graphene based devices. (a) transfer characteristics of device D1 measured at different temperatures.	(b) Mobility extracted from   Fig.~\ref{Fig2}a    curve using the Eq.~(\ref{mu}) at room temperature. (c) Scattering time extracted from the Fig.~\ref{Fig2}a curve using the Eq.~(\ref{sigma}) at room temperature. } \label{Fig2}
\end{figure}

We extract mobility and transport scattering time from the conductance measurements shown in Fig.~\ref{Fig2}a. The conductivity of graphene is given by
\begin{equation}
\sigma = \frac{e^2}{\pi \hbar  }  \left[\frac{E \tau (E)}{\hbar} \right]_{E=E_{\rm F}}, \label{sigma}
\end{equation}
where $E_{\rm F}$ is the Fermi energy and  $\tau(E)$ is the energy-dependent  transport scattering time. The Fermi energy depends on the concentration,  $E_{\rm F}= \hbar  v_{\rm F} \sqrt{\pi n},$
which is controlled by the gate voltage:  $n=\varepsilon U_{\rm g}/4\pi e d,$ where $U_{\rm g}$ is the gate voltage counted from the Dirac point,   $\varepsilon$ is the dielectric constant and $d$ is the width of the spacer.  Here, in all calculations,  we assume that $U_{\rm g}>0$ having in mind  that response should change sign under replacement $U_{\rm g} \rightarrow -U_{\rm g}. $ Using Eq.~(\ref{sigma}) and the formula
\begin{equation}
E_{\rm F}= \hbar v_{\rm F} \left(\frac{\varepsilon U_{\rm g}}{4 e d}\right)^{1/2},
\end{equation}
one can extract from experimental data (see Fig.~\ref{Fig2}a)  the  dependence of both   $\tau$ and mobility, defined as
\begin{equation} 
\mu= \frac{1}{e} \frac{d \sigma}{ dn }=  \frac{1}{e} \frac{d \sigma}{dU_{\rm g} }\frac{d U_{\rm g}}{ dn }, \label{mu}
	\end{equation}
on $U_{\rm g}.$ The result is presented in Figs.~\ref{Fig2}b and c. Assuming that with approaching to the neutrality  point the  scattering is limited by charged impurities, for which
\begin{equation}
\gamma=  \tau^{-1} \propto 1/E \propto 1/\sqrt{U_{\rm g}},
\label{gamma}
\end{equation}
we approximate the experimental dependence  shown in Fig.~\ref{Fig2}b as
\begin{equation}
\tau(U_{\rm g})=\tau^*~\frac{\sqrt{U_{\rm g}}}{\sqrt{U_{\rm g}} + \sqrt{U_*}}, \label{tauUg}
\end{equation}
where  $\tau_* = 2 \cdot 10^{-14}$s   and $U_* =1.5 V.$ Equation~(\ref{tauUg}) was used to calculate $\Omega$ and $\Gamma$ entering Eq.~(\ref{OmegaGamma}).  We also  used  the well known  expression for plasma wave velocity in graphene
\begin{equation}
s=\left( \frac{4 e^3 d  v_F^2 U_{\rm g}}{\varepsilon \hbar^2 } \right)^{1/4}. \label{sUg}
\end{equation}
Equations~(\ref{tauUg}) and (\ref{sUg}) become invalid very close to the neutrality point.  Finally, we note that the $G(U_{\rm g})$ curves do not change as the temperature goes down, meaning that mobility is restricted by the defect scattering. 

\textbf{Laser and experimental setup. } The experiments have been performed applying a continuous wave (CW) methanol laser operating at frequencies $f_1 = 2.54$~THz (wavelength $\lambda_1 = 118$~$\mu$m) with a power of $P \approx 20$~mW and $f_2 = 0.69$~THz (wavelength $\lambda_2 = 432$~$\mu$m) with $P \approx 2$~mW \cite{Ganichev2009, Kvon2008}. The incident radiation power was monitored by a reference pyroelectric detector. The laser beam was focused onto the sample by a parabolic mirror with a focal length of 75~mm. Typical laser spot diameters varied, depending on the wavelength, from 1 to 3~mm. The spatial laser beam distribution had an almost Gaussian profile, checked with a pyroelectric camera \cite{Ziemann2000}. The radiation was modulated at about 75~Hz  by an optical chopper. The samples were placed in an optical temperature variable cryostat and photoresponse was measured as a voltage drop using standard lock-in technique. The radiation of the laser was linearly polarized. To demonstrate sensitivity of the photovoltage to the helicity of the incoming radiation we place a $\lambda/4$ plate made of $x$-cut crystalline quartz in the incoming beam. The helicity of the radiation is then controlled via changing the angle $\phi$ between the laser polarization and the main axes (``fast direction'') rotating the plate in vertical plane. When $\varphi = 45^\circ$ the radiation incident on the device is clock wise circularly polarized (right circularly polarized radiation, $\sigma^+$) while for $\phi = 135^\circ$ one gets the opposite polarization (left circularly polarized radiation, $\sigma^-$). To study response to the linearly polarized radiation we used $\lambda/2$-plates. Rotating the plate, we rotated the radiation electric field vector $\mathbf{E}$ by an azimuth angle $\alpha$. 

\textbf{Fitting parameters.} The curves in Fig.~\ref{fig:helicity} were obtained with fitting parameters summarized in Tab.~\ref{table1}.

\begin{table}
	\begin{tabular}{lrrrrrr}
		\hline
	Panel in Fig.~\ref{fig:helicity} 	& a)     & b)    & c)    & d)     & e)    & f)   \\
		\hline
		$U_{\rm C}/P$  & -0.012 & -0.42 & 1.15  & 0.009  & 0.15  & 2.15 \\
		$U_{\rm L1}/P$ & 0.025  & -0.14 & -0.08 & -0.007 & -0.09 & 3.34 \\
		$U_{\rm L2}/P$ & -0.039 & 0.49  & -1.03 & -0.009 & 0.01  & 0.52 \\
		$U_0/P$  & -0.005 & 0.46  & 3.18  & 0.003  & 0.13  & 2.58 \\
		\hline
	\end{tabular}
		\caption{Fitting parameters used for calculations of curves in Fig.~\ref{fig:helicity}. The parameters are given in mV/W. Letters on the top of the table corresponds to the labels of panels in Fig.~\ref{fig:helicity}. }
		\label{table1}
\end{table}

\subsection{ Response to polarization of arbitrary polarization}

We assume that wavelength of the radiation is much larger that the device size and that  incoming beam linearly polarized along $x$ axis acquires elliptical polarization after transmission through $\lambda/4$ plate. Then, the field is described by homogeneous electric vector $\mathbf E(t) $ with the components
\begin{eqnarray}\label{Ex}
  E_x(t) &=&\frac{\left( E_x e^{-i\omega t} +  \rm{h.c.} \right)}{2} =E_0\cos\alpha \cos(\omega t),  \\
  \label{Ey}
 E_y(t) &=&\frac{\left( E_y e^{-i\omega t} +  \rm{h.c.} \right)}{2} =E_0\sin\alpha \cos(\omega t+\eta)
\end{eqnarray}
where
\begin{equation}
E_x=E_0\cos \alpha,\quad E_y=  E_0 \sin \alpha e^{-i \eta}.
\label{Exy1}
\end{equation}
 We get
 \begin{eqnarray}\label{Stokes0}
    |E_x|^2+ |E_y|^2&=&E_0^2,  \\
\label{StokestildeL}
   E_x E_y^* + E_x^* E_y    &=& E_0^2 ~  P_{\rm L1}, \\
    \label{StokesL}
   |E_x|^2- |E_y|^2 & = &E_0^2 ~P_{\rm L2},\\  \label{StokesC}
  i\left( E_x E_y^* - E_x^* E_y \right)  &=& - E_0^2 ~P_{\rm C}.
 \end{eqnarray}
\begin{figure}[b]
\centerline{\includegraphics[width=0.8\columnwidth]{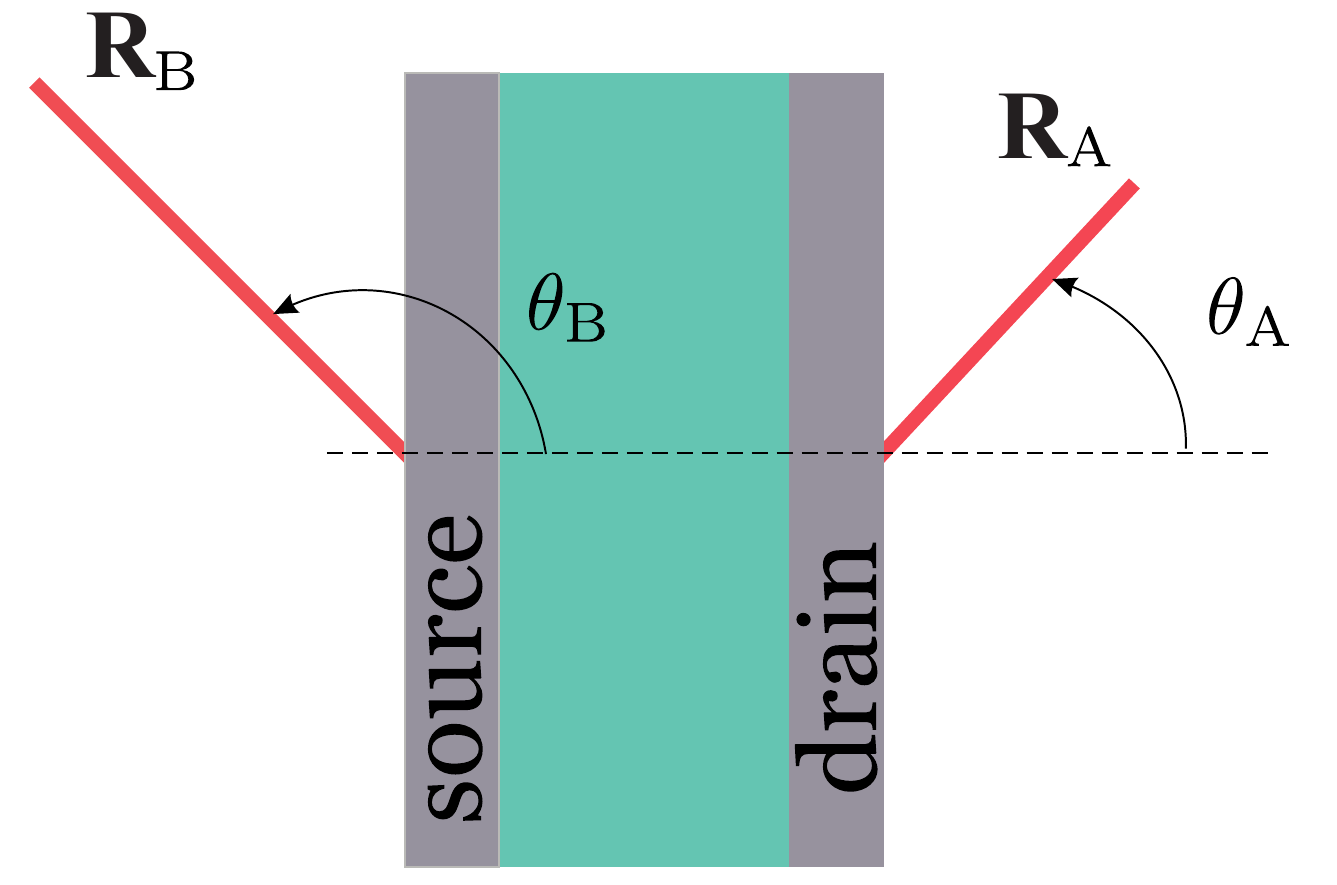}}
\caption{(Color online) Sketch of the  FET with two antennas}
\label{Fig1}
\end{figure}
Here, we introduced  Stokes parameters
\begin{eqnarray}
\label{Stokes1tildeL}
    P_{\rm L1} = & \sin(2\alpha) \cos \eta  = \frac{\sin(4\phi)}{2}, \\
\label{Stokes1L}
      P_{\rm L2} = & \cos(2\alpha)  =  \frac{1+\cos(4 \phi)}{2},\\
   P_{\rm C}  =  & \sin(2\alpha)\sin\eta  =  \sin (2 \phi), 
   \label{Stokes1C}
 \end{eqnarray}
where $\phi$ is the rotation angle of $\lambda/4$ plate with respect  to $x$ axis. As seen, the angles $\phi=0, \pm 90^\circ,\pm 180^\circ$ correspond to linear polarization along $x-$axis, while angles $\phi=45^\circ$ (or $\phi=225^\circ$) and  $\phi=135^\circ$ (or $\phi=315^\circ$ ) to circular polarization with right and left helicity, respectively. We note  that the helicity  also changes sign  by inversion of the radiation frequency:   $\omega \to -\omega.$  For definiteness, in all equations below  we  assume $\omega>0.$

The detailed analysis of  time-dependent potential distribution in our device requires calculation of antennas  properties, which is a challenging problem beyond the scope of the current work. However, some phenomenological properties of suggested interferometer can be understood by using  a toy model, which captures basic physics of the problem. The model is illustrated in Fig.~\ref{Fig1}, where two antennas are replaced with long thin metallic  rods described by vectors $\mathbf R_{\rm A,B}$.  Assuming that antennas
are perfect conductors and    neglecting small mutual capacitances,   one
can  write the potentials applied to source and drain as
\begin{eqnarray} & \nonumber U_{\rm A}(t) = \mathbf E(t) \mathbf R_{\rm A}/2 =U_{\rm A}
\cos (\omega t -\varphi_{\rm A})  ,   \\
& \nonumber U_{\rm B}(t) = \bm E(t) \mathbf R_{\rm B} /2=U_{\rm B}\cos (\omega t -\varphi_{\rm B}),    \end{eqnarray}
where parameters $U_{\rm A, B}$ and $\varphi_{\rm A, B}$  for $\omega>0$ obey  [see
Eqs.~(\ref{Ex}) and (\ref{Ey})]
\begin{eqnarray} \label{A}
& U_{\rm A} e^{-i\varphi_{\rm A}}= \frac{E_0 R_A}{2}\left( \cos \alpha \cos \theta_{\rm A}
+ \sin \alpha \sin \theta_{\rm A}~ e^{-i \eta} \right), \\
& U_{\rm B} e^{-i\varphi_{\rm B}}= \frac{E_0 R_{\rm B}}{2}\left( \cos \alpha \cos \theta_{\rm B} + \sin \alpha \sin \theta_{\rm B} ~ e^{-i \eta} \right).
\label{B}
\end{eqnarray}

In the simplest case of circular polarization considered in Refs.~\cite{S1,S2}($\alpha= \pm 45^\circ,  \eta= 90^\circ$), equations (\ref{A}) and (\ref{B}) simplifies, so that $U_{\rm A,B}=E_0 R_{\rm A,B}/\sqrt{8} $  and $\varphi_{\rm A,B}=\theta_{\rm A,B}$ for $\omega>0$ (positive helicity) and $\varphi_{\rm A,B}=-\theta_{\rm A,B}$   for $\omega<0$ (negative helicity).  For the case of generic elliptic  polarization, it is convenient to use the Stokes parameters. After simple algebra, we find equations needed for calculation of the response according to the theory developed in Refs.~\cite{S1,S2}
$$ 
U_{\rm A}^2-U_{\rm B}^2  =  a_0  +  a_{\rm L1}   P_{\rm L1}+ a_{\rm L2} P_{\rm L2} =
$$
\begin{equation} 
\label{UA} 
 = a_0 + a_{\rm L2} \frac{1+\cos(4 \phi)}{2} +  a_{\rm L1}  \frac{\sin(4\phi)}{2}, 
\end{equation}
\begin{equation}
\label{UB} 
U_{\rm A} U_{\rm B} \sin(\phi_{\rm A} -\phi_{\rm B}) = a_{\rm C} P_{\rm C}
= a_{\rm C} \sin(2 \phi),
\end{equation}
where
\begin{eqnarray}
\label{J0}
  a_0 &= \frac{E_0^2(R_{\rm A}^2-R_{\rm B}^2)}{2},
  \\
  \label{tildeJL}
   a_{\rm L1} & = \frac{E_0^2
  \left[R_{\rm A}^2\sin(2 \theta_{\rm A}) -R_{\rm B}^2\sin( 2 \theta_{\rm B} \right]}{2},
  \\
  \label{JL}
  a_{\rm L2 }& =\frac{E_0^2 \left[R_{\rm A}^2\cos(2 \theta_{\rm A})
   -R_{\rm B}^2\cos( 2 \theta_{\rm B}) \right]}{2},
   \\
    \label{JC}
  a_{\rm C} & =-\frac{E_0^2 R_{\rm A} R_{\rm B}}{2} \sin(\theta_{\rm A} -\theta_{\rm B})
\end{eqnarray}
are purely geometrical factors.  Although these factors can change for a more realistic models of antennas,  the general structure of Eqs.~\ref{UA} and \ref{UB}  should remain the same, so that one can use $a_0, a_{\rm L1 },  a_{\rm L2 }$ and   $a_{\rm C} $ as fitting parameters that do not depend on frequency and gate voltage.   Importantly,  terms proportional  to $U_{\rm A}^2- U_{\rm B}^2$ and to  $U_{\rm A} U_{\rm B}$ have different  angle periodicity---$\pi/2$ and $\pi,$ respectively---that allows one to separate them in experiment. Dependence of the response $U$ on $\phi$ can be found by using results of Ref.~\cite{S1} [see Eqs.~(14-16) in  \cite{S1}] :
 \begin{equation}
  U(\phi)= \frac{\omega}{\sqrt{\omega^2 +\gamma^2}}\frac{\alpha (U_{\rm A}^2-U_{\rm B}^2)
  +\beta U_{\rm A} U_{\rm B} \sin (\theta_{\rm A} -\theta_{\rm B})}{4 U_g \left| \sin (kL)\right|^2} ,
  \label{response1}
  \end{equation}
$$
  =  \frac{\omega}{4 U_{\rm g} \left| \sin (kL)\right|^2\sqrt{\omega^2 +\gamma^2}}  \left\{ \displaystyle \alpha \left[a_0 + a_{\rm L1} \frac{\sin (4 \phi)}{2} \right. \right. + 
$$
  \begin{equation}
  + \left. \left. a_{\rm L2}\frac{1+ \cos(4\phi)}{4}    \right]+   \beta a_{\rm C} \sin(2 \phi)\right \},
\label{response2}
  \end{equation}
where $s$ is the plasma wave velocity, $L$ is the length of the FET channel,
  \begin{equation} k=  \frac{\sqrt{\omega(\omega+ i \gamma)}}{s} =\frac{\Omega+ i \Gamma}{ s},\end{equation}
is the plasma wave vector, $\gamma$ is the inverse momentum relaxaton   time, and $\Omega$ and  $\Gamma$
are effective frequency and damping rate of the plasma waves, given by
  \begin{equation}
 \label{OmegaGamma} \Omega = \sqrt{\frac{\sqrt{\omega^4+ \omega^2 \gamma^2} +\omega^2 }{2}} ,
 \quad \Gamma =  \sqrt{\frac{\sqrt{\omega^4+ \omega^2 \gamma^2} -\omega^2 }{2}} .
\end{equation}
and the coefficients $\alpha$ and $\beta$ read
  \begin{eqnarray} \label{alpha}
  & \alpha= \left(\! 1\!+\!\frac{\gamma \Omega\!}{ \Gamma \omega}\right) \sinh^2\!\left(\!\frac{\Gamma L}{s}\!\right)\!-\! \left( \! 1\!-\!\frac{\Gamma \gamma}{ \Omega \omega}\!\right) \sin^2\!\left(\!\frac{\Omega L}{s}\!\right),
  \\
  & \beta=8\sinh\!\left(\!\frac{\Gamma L}{s}\!\right)\sin\!\left(\!\frac{\Omega L}{s}\!\right).
    \label{beta}
    \end{eqnarray}
Equation~(\ref{response2}) can be written in form of Eq.~(\ref{eqn:response2}) with
\begin{eqnarray}\label{F1}
   &F_1=  \frac{ \omega~ \alpha }{4 U_{\rm g} \left| \sin (kL)\right|^2\sqrt{\omega^2 +\gamma^2}},   \\
   &F_2=\frac{\omega ~\beta}{4 U_{\rm g} \left| \sin (kL)\right|^2\sqrt{\omega^2 +\gamma^2}}.
\label{F_2}
\end{eqnarray}
The coefficients $U_0,~U_{\rm L1},~U_{\rm L2},$ and $U_{\rm C}$ entering Eq.~(\ref{eqn:eq1}) reads
\begin{equation}
U_0= F_1 a_0 , ~U_{\rm L1}= F_1 a_{\rm L 1},~ U_{\rm L2}= F_1 a_{\rm L2},~  U_{\rm C}= F_2 a_{\rm C}.
\end{equation}

\vspace{1mm}

\subsection{Interference part of the response}
Eq.~(\ref{response2}) allows to extract experimentally  interference contribution to  response.   We find
$$
U_{\rm C} = \frac{U(\phi = 45^\circ) -U(\phi = 135^\circ)}{2}= \frac{U_{\sigma^+} -U_{\sigma^-} }{2}= 
$$
\begin{equation}
\label{interference}
= \frac{\displaystyle  2\omega  a_{\rm C}\sinh\!
\left({\Gamma L}/{s}\!\right)\sin\!\left({\Omega L}/{s}\right)}{ U_{\rm g} \left| \sin (kL)\right|^2
\sqrt{\omega^2 +\gamma^2}}.
\end{equation}
Hence, interference term in our  setup depends only  on   circular component of polarization.  Using simplified formula  for geometrical factor, $J_{\rm C},$   [see Eq.~(\ref{JC})],  we rewrite Eq.~(\ref{interference}) as follows
$$
U_{\rm C}=  \frac{E_0^2 R_{\rm A } R_{\rm B}}{4 U_{\rm g}} \frac{\omega}{\sqrt{\omega^2 +\gamma^2} }  \times
$$
\begin{equation} 
\label{final}
\times \frac{\displaystyle  \sinh\! \left({\Gamma L}/{s}\!\right)\sin\!\left({\Omega L}/{s}\right) \sin(\theta_{\rm A} -\theta_{\rm B} )}{ \left| \sin (kL)\right|^2}.
\end{equation}
We see that the response is proportional to
$\sin(\theta_{\rm A} -\theta_{\rm B} ).$
One can show that for device with a  beam  splitter and  a delay line one should simply add in Eq.~(\ref{final}) the delay line  phase shift $\delta$ to geometrical phase shift
$\theta_{\rm A} -\theta_{\rm B}$:
\begin{equation}
\sin(\theta_{\rm A} -\theta_{\rm B} ) \to  \sin(\theta_{\rm A} -\theta_{\rm B} + \delta  ).
\end{equation}


\subsection{Non-resonant regime}
Equations (\ref{OmegaGamma}), (\ref{alpha}), (\ref{beta}), (\ref{F1}) and (\ref{F_2})  simplify in the non-resonant regime,
$\omega \ll \gamma, ~ s/L \ll \gamma$  under additional assumption $L \gg  L_*$ (long non-resonant  sample \cite{S1}).  Here 
\begin{equation}
    L_*= \frac{s\sqrt 2}{\sqrt {\omega \gamma}}
    \label{eqn:L_star}
\end{equation}
is the plasma wave decay length. In this regime, plasma waves, excited at the source and drain parts of the channel weakly overlap  inside the transistor  channel. The  response  becomes  \cite{S1,b3}%
$$
U=\frac{U_{\rm A}^2- U_{\rm B}^2}{4 U_{\rm g}} +   
$$
\begin{equation}
+ \frac{16 U_{\rm A} U_{\rm B} e^{-L/L_*} \sin(L/L_*) (\omega/\gamma) \sin (\varphi_{\rm A}-\varphi_{\rm B}) }{4 U_{\rm g}},   
\label{eqn:simple_eq}
\end{equation}
while the  coefficients  $F_1$ and $F_2$ are given by  Eqs.~(\ref{F12}) of the main text. Physically, last phase- and helicity-sensitive   term in Eq.~(\ref{eqn:simple_eq}) describes interference  arising due to overlapping  of the plasma waves.   


	

\begin{thebibliography}{100}
\bibitem{b2} From the experimental conductance curves  the scattering rate $\gamma$ is estimated to be about 50~THz, which is much larger than the radiation frequency 2.54~THz used in our work.
\end{thebibliography}

\end{document}